\documentclass[useAMS,usenatbib]{mn2e}
\usepackage{graphicx}
\usepackage{amssymb}
\usepackage{lscape}
\usepackage{ulem}
\usepackage{txfonts}

\def\kms{km~s$^{-1}$}

\def\apjs{ApJS}

\def\msun{M_{\odot}}

\title[The orbital extent of enhanced star formation]
{Galaxy pairs in the Sloan Digital Sky Survey - VI. 
The orbital extent of enhanced star formation in interacting galaxies}

\author[Patton et al.] {David R. Patton$^1$, Paul Torrey$^2$, Sara L. Ellison$^3$, 
J. Trevor Mendel$^3$ and Jillian M. Scudder$^3$\\ 
$^1$ Department of Physics and Astronomy, Trent University, 
1600 West Bank Drive, Peterborough, ON K9J 7B8, Canada.\\
$^2$ Harvard Smithsonian Center for Astrophysics, 60 Garden Street, Cambridge, MA 02138, USA\\
$^3$ Department of Physics and Astronomy, University of Victoria, Victoria, BC  
V8P 1A1, Canada\\
}

\begin{document}
\maketitle

\begin{abstract}
We use pair and environmental classifications of $\sim$ 211 000 star-forming galaxies 
from the Sloan Digital Sky Survey, along 
with a suite of merger simulations, to investigate the enhancement of star formation as a 
function of separation in galaxy pairs.  Using a new technique for distinguishing between the 
influence of nearby neighbours and larger scale environment, we find a clear enhancement in 
star formation out to projected separations of $\sim$ 150 kpc, beyond which there is no 
net enhancement.  We find the strongest enhancements at the smallest separations 
(especially $< 20$ kpc), consistent with earlier work.  Similar trends are seen in 
the simulations, which indicate that the strongest enhancements are  
produced in highly disturbed systems approaching final coalescence, whereas the 
more modest enhancements seen at wider separations are the result of 
starburst activity triggered at first pericentre passage, 
which persists as the galaxies move to larger separations.
The absence of any net enhancement beyond 150 kpc provides reassurance that the detected 
enhancements are due to galaxy$-$galaxy interactions, rather than larger scale environmental 
effects or potential pair selection biases.
A rough census indicates that 66 per cent of the enhanced star formation in our pair sample occurs at 
separations $>$ 30 kpc.  We conclude that significant interaction-induced 
star formation is not restricted to merger remnants or galaxies with close companions; 
instead, a larger population of wider separation pairs exhibit enhanced star formation 
due to recent close encounters.
\end{abstract}


\section{Introduction}
The observational study of close galaxy pairs has revealed clear evidence of enhanced 
star formation rates (SFRs) in interacting and merging galaxies.  
The strongest enhancements are found in the closest pairs, with projected separations 
$\lesssim 30$ kpc \citep[e.g.,][]{barton00,lambas03,ellison08,woods10}.  
These enhancements are consistent with interaction-induced star formation seen in 
simulations of merging galaxies, 
with the closest pairs comprised of systems seen near their first pericentre passage, 
as well as those which are coalescing \citep[e.g.,][]{mihos96,dimatteo07,cox08,montuori10}.

More recently, \citet{scudder12} have shown that SFR enhancements 
are present out to the 80 kpc limit of their pairs sample.  These enhancements 
are accompanied by bluer central colours \citep{patton11}, 
diluted metallicities \citep{scudder12}, 
and an increased incidence of active galactic nuclei \citep[AGNs;][]{ellison11} 
and luminous infrared galaxies \citep{ellison13}. 
While some studies have probed enhanced star formation out to even larger separations, 
it is unclear if these enhancements are due to galaxy$-$galaxy interactions, 
larger scale environmental effects or sample biases \citep[e.g.,][]{alonso06,park07,robaina09,robaina12}.  

This leaves open a crucial question: at what projected separations do interaction-driven  
enhancements disappear?  The answer to this question is needed for a full 
accounting of the contributions of interaction-induced star formation to the cosmic SFR, 
and can also provide insight into the orbits of interacting galaxies.
The orbital extent of enhanced star formation depends on a number of factors, such as 
the energies and angular momenta of pair orbits, the strength and duration of 
induced star formation, the availability and distribution of gas, stellar and AGN feedback, 
the amount of contamination by non-interacting pairs (e.g., chance superpositions 
within groups or clusters), etc.  

With this Letter, we address this question using a spectroscopic 
sample of Sloan Digital Sky Survey (SDSS)\footnote{sdss.org} galaxy pairs which extends out 
to much larger separations than earlier work, and which carefully controls for 
environment.  We then compare our findings 
with predictions from a suite of merger simulations spanning a range of orbital eccentricities, 
impact parameters and disc orientations.
We adopt a concordance cosmology of $\Omega_{\Lambda} = 0.7$, $\Omega_{\rm M} = 0.3$ 
and $H_0 = 70~h_{70}$ \kms Mpc$^{-1}$ throughout.

\section{Sample Selection}\label{secsample}

\subsection{Input catalogue}

We begin by identifying a large spectroscopic sample of galaxies drawn from the 
SDSS Data Release 7 \citep[DR7;][]{dr7}.  We require that 
all galaxies have secure spectroscopic redshifts (zConf $>$ 0.7), redshifts 
of $0.02 < z < 0.2$, 
photometric and spectroscopic classifications as galaxies, 
and extinction-corrected Petrosian apparent magnitudes of 
$14.0 \le m_r \le 17.77$.

We employ the photometric total stellar mass measurements of \citet{mendel13}, which were 
estimated using the updated $g-$ and $r-$band photometry of \citet{simard11}, along with new 
photometry in the $u-$ and $i-$bands.  
We note that the photometry of \citet{simard11}
has been shown to provide a significant improvement over the standard SDSS pipeline, especially  
for galaxies in crowded systems such as close pairs \citep{patton11}.  Following \citet{patton11}, 
we take full 
advantage of this improved photometry by requiring that each galaxy's observed 
fibre $g-r$ colour be within 0.1 mag of its \citet{simard11} model-predicted fibre colours.    
This yields an input catalogue of $\sim$ 607 000 
galaxies for which secure redshifts and stellar masses are available.

\subsection{Pair and environmental classifications}

For every galaxy in this catalogue, we identify the closest neighbour, and 
characterize the environment with several additional measurements.  We define the 
closest neighbour to be the galaxy with the smallest projected physical separation 
(hereafter $r_p$) which has a rest-frame relative velocity (hereafter $\Delta v$) 
of $<~1000$~\kms~ and is within a factor of 10 of the galaxy's stellar mass.
We also record the projected separation of the second closest neighbour (hereafter $r_2$) and the 
total number of neighbours within 2 Mpc (hereafter $N_2$).

\subsection{Star formation rate measurements}

We now focus on galaxies which have secure SFR measurements, in order to examine 
how the presence of nearby companions (with or without SFR measurements) affects galactic 
star$-$forming properties.
Our SFR measurements are drawn from the catalogue 
of \citet{brinchmann04}, which has been updated to include DR7.  Following \citet{scudder12}, 
we restrict our analysis to fibre SFRs only.  
We require each galaxy to be classified as star-forming 
using the criteria of \citet{kauffmann03}, 
and we require a signal$-$to$-$noise ratio of at least one in each of the emission lines 
used for this classification.
These criteria yield a sample of $\sim$ 211 000 star-forming galaxies.  
No SFR requirements are imposed on neighbouring galaxies; 
therefore, our sample will include star forming members of 
both `wet' and `mixed' galaxy pairs.

\subsection{Control sample}

We wish to determine the influence (if any) that nearby companions have on the star$-$forming 
properties of galaxies.  A common approach is to compare with a control sample of galaxies 
which have similar selection criteria but no close companions 
\citep[e.g.,][]{ellison08,patton11,scudder12}.  However, this approach is robust only 
for relatively close companions; extending this approach to wider separations yields 
a control sample whose galaxies are substantially more isolated than the paired galaxies 
they are being compared with.  

One method of dealing with this issue is to match paired and control galaxies on 
local density \citep{ellison10,scudder12}.  However, 
as local density probes a substantially larger scale than the pairs, 
this matching cannot account for important differences in the smaller scale environment, 
as exemplified by compact groups embedded in loose groups \citep{mendel11}.  
Another approach is to require that both pairs and 
control galaxies be isolated on larger scales \citep{barton07}.  However, 
as most galaxies reside in groups or clusters, this may remove the majority of the 
sample under consideration.  Instead, we introduce a new methodology which addresses this 
issue by matching the control sample in both local density and isolation.  A brief 
description of this method is provided here, with additional details and validation 
deferred to a subsequent paper (Patton et al. in preparation).

For each galaxy, we identify a control sample which is matched
in stellar mass, redshift, local density, and isolation.  We begin by requiring  
each control galaxy to be within 0.01 in redshift and 0.1 dex in stellar mass.  
We then use $N_2$ as a proxy for local density (which is reasonable given the match 
in redshift and stellar mass), and require that the $N_2$ of a galaxy and its controls agree 
within 10 percent.  Finally, to match in isolation, we require the $r_p$ of the 
control galaxy's closest companion to be within 10 per cent of the distance to the 
galaxy's second closest companion ($r_2$).   
Control galaxies are chosen with replacement, yielding an average of 34 controls for each galaxy. 
We determine weighted mean properties of each galaxy's  
control sample, assigning higher statistical weights to the control galaxies which 
provide the best simultaneous matches in $z$, mass, $N_2$ and $r_2$.  This approach 
yields excellent agreement between galaxies and their controls in the four properties 
that are being matched.  The same statistical weights are then also used to compute 
the mean SFR of the controls for each galaxy.

\section{Enhanced Star Formation in SDSS Pairs}\label{sdss}

Armed with the SFRs of galaxies and their statistical control samples, we now 
proceed with a direct comparison between the two, as a function of pair separation.  First, 
following \citet{scudder12}, we restrict our analysis to pairs with $\Delta v < 300$ \kms, 
thereby focusing on the pairs which are most likely to be undergoing interactions.
Next, we must account for the under-selection of close angular pairs ($< 55$ arcsec) 
in SDSS due to fibre collisions.  \citet{ellison08} dealt with this effect by randomly 
culling 67.5 per cent of pairs with separations $> 55$ arcsec, using the results of the 
\citet{patton08} census of SDSS pairs.  
We instead apply a statistical weight $w_{\theta}$ (as defined in \citealt{patton02}) 
of 3.08 to each galaxy in close angular pairs ($w_{\theta} = 1/(1-0.675)$), 
thereby avoiding a cull of the wide pairs crucial to this analysis.

\begin{figure}
\centerline{\rotatebox{0}{\resizebox{8.0cm}{!}
{\includegraphics{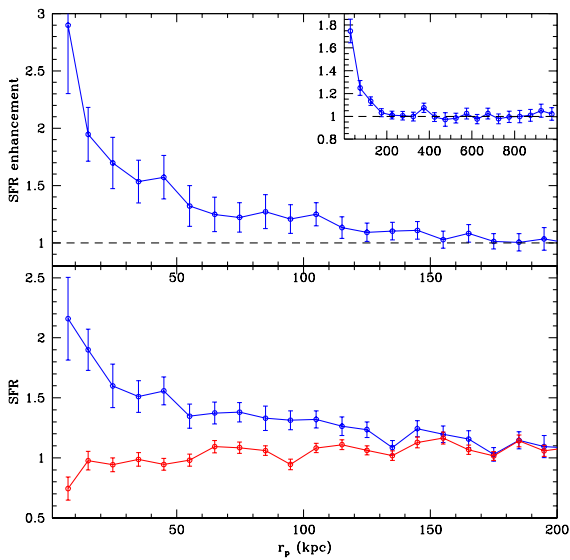}}}}
\caption{
Lower panel: the mean SFR of SDSS paired galaxies (blue symbols) and their associated 
control galaxies (red symbols) is plotted versus projection separation ($r_p$).
Upper panel: mean SFR enhancement (ratio of pair SFR to control SFR) is plotted versus 
$r_p$, with the dashed horizontal line denoting zero enhancement.
The inset plot extends this out to 1000 kpc, using larger $r_p$ bins.
All error bars show the standard error in the mean.
\label{figrpsfr}}
\end{figure}

\begin{figure}
\centerline{\rotatebox{90}{\resizebox{5.4cm}{!}
{\includegraphics{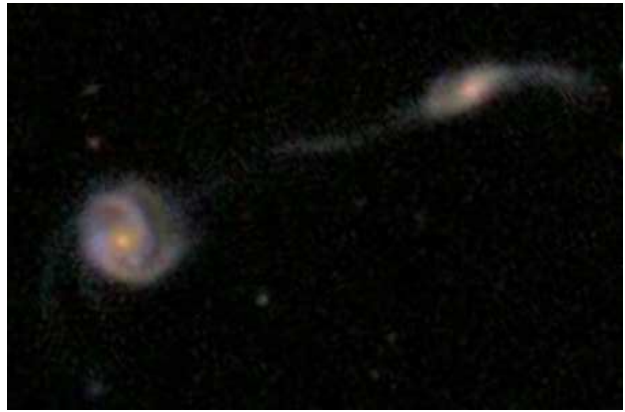}}}}
\caption{This SDSS $gri$ image shows a wide galaxy pair ($r_p = 91$ kpc) 
with an obvious tidal tail linking the member galaxies.  
Tidal features are also seen on the opposite side of each galaxy.  
The galaxy on the right (SDSS objID=587736619324735891) 
has an SFR of 7.8 $\msun$~yr$^{-1}$, which is 5.6 times higher than its controls.
The companion galaxy is not in our star-forming sample, 
since it has a composite spectrum (star formation + AGN).
\label{figtidaltail}}
\end{figure}

In the lower panel of Fig.~\ref{figrpsfr}, we plot the mean SFR of paired and 
control galaxies as a function of $r_p$.  In the upper panel of the same figure, 
we plot the ratio of pairs versus control SFRs, which we interpret as the enhancement 
of SFR due to interactions (a ratio of one means no enhancement).  
At small separations, paired galaxies are 
found to have substantially higher SFRs than controls (up to a factor of 3), 
with very high statistical 
significance.  Pair SFRs exhibit moderate enhancements of $\sim$ 20 per cent  
over the range $50 < r_p < 120$ kpc,
and then decline to match their controls by $r_p \sim 150$ kpc.  
Fig.~\ref{figtidaltail} shows a striking example of enhanced star formation 
in a relatively wide pair (91 kpc) 
with a clear tidal tail linking the member galaxies.  
Above 150 kpc, 
there is no evidence of net enhancement (or suppression) in the pair SFR  
(this is true out to at least 1000 kpc, as shown in the inset 
panel of Fig.~\ref{figrpsfr}).  

Over the range $r_p < 80$ kpc, these results are broadly consistent 
with the findings of \citet{scudder12}, despite some differences in data (e.g., stellar masses, 
S/N requirements), pair selection, and environmental matching.  
At larger separations, our results indicate that we have achieved our goal of measuring 
SFR enhancements out to sufficiently wide separations that the enhancements reach zero.  
Since our paired galaxies have controls which are closely matched in galaxy properties 
(stellar mass and $z$) and environment (local density and isolation), 
we conclude that galaxy$-$galaxy interactions appear to be able to boost 
the mean SFR out to $r_p \sim 150$ kpc, but not further.  
If true, this implies that enhanced star formation extends out to larger 
separations than has previously been appreciated, and that studies of strongly 
interacting galaxies may miss a sizeable population of galaxies 
which exhibit enhanced star formation due to recent close encounters.  
To examine if this interpretation is physically plausible, we now turn to simulations.

\section{$N-$BODY/SMOOTHED PARTICLE HYDRODYNAMICS (SPH) SIMULATIONS}

\subsection{Overview of the Simulations}

We investigate the possibility that mergers are 
responsible for the observed SFR enhancement using a controlled suite of numerical
simulations.  The numerical methods employed here are similar to those described in detail 
in~\citet{torrey12a}.    Specifically, our simulation suite has been run using 
the N-Body/SPH simulation code {\small GADGET}~\citep{gadget} which -- in addition 
to including gravity and hydrodynamics -- accounts for radiative cooling of 
gas~\citep{KatzCooling}, a density$-$driven SFR prescription with 
associated feedback~\citep{SHMultiPhase}, 
and gas recycling from aging stellar populations~\citep{torrey12a}.  

Our goal is to construct a set of galaxy merger simulations, where 
we can compare the evolving SFRs in isolated galaxies to the SFRs 
in merging systems, as a function of galactic separation $r$.  
To achieve this, we construct two isolated galaxies of initial stellar mass 
$M_1=5.7\times10^{9} {\rm M}_\odot$ and
$M_2 = 1.4\times 10^{10} {\rm M}_\odot$.  These masses, and the resulting
stellar mass ratio of $\sim 2.5:1$, were chosen to match the median mass and 
mass ratio of our SDSS pairs sample. 
Each isolated galaxy contains a dark matter halo, a stellar and gaseous disc  
(20 per cent initial gas fraction), and a stellar bulge (which contains 
20 per cent of the stellar mass).
We ensure that our galaxies are stable against bar formation or other instabilities 
when evolved in isolation, and then set them on merging trajectories 
as described in the following subsection.
Since our isolated galaxies are stable against bar formation or any rapid 
changes in their SFR, any major changes seen in the 
SFR for the merging galaxies can be confidently attributed to
the merger interaction.

\subsection{A suite of 75 merger simulations}

For our merger suite, we adopt a set of five eccentricities (0.85, 0.9, 0.95, 1 and 1.05)
and five impact parameters (2, 4, 8, 12 and 16 kpc) which are consistent with 
orbital element distribution functions derived from cosmological simulations~\citep{wetzel2011}. 
While many previous merger studies have limited their scope to studying zero energy orbits
with fixed impact parameters,
the 25 combinations of eccentricity and impact parameter used in our merger suite 
yields substantial differences in the resulting apocentre separations ($r_{\rm apo}$).  
This is critical for this Letter, where we are interested in assessing 
the feasibility of driving SFR enhancements at relatively large galactic separations 
($> 100$ kpc) via galaxy$-$galaxy interactions.  Since there are no known 
correlations between the orientation of each galaxies' angular momentum vector 
relative to the plane of the merger~\citep[e.g.][]{Khochfar2006}, we select 
three representative merger disc orientations 
(the e, f and k orientations from \citealt{robertson06}), yielding a suite 
of 75 merger simulations.  We emphasize that this merger suite is not intended to be complete,  
as we are still failing to sample a large portion of available merger parameter space (e.g., 
variations in galaxy morphology, mass, mass ratio, gas fraction, etc.).  Moreover, these 
simulated pairs are pure binary systems, without the cosmological context of additional 
galaxies spanning a range of environments.  
However, by including 
a range of orbital parameters as we have done here, we gain specific insight about 
interaction-driven starburst activity at large galactic separations which has not been 
previously extensively studied.

\begin{figure}
\centerline{\rotatebox{0}{\resizebox{8.0cm}{!}
{\includegraphics{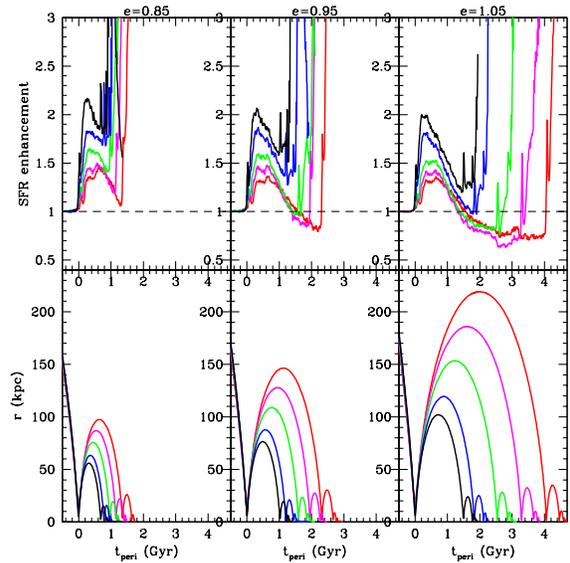}}}}
\caption{A subset of our suite of simulated orbits is shown.  The lower panels depict 3D  
pair separation ($r$) versus time since first pericentre ($t_{\rm peri}$).  
The upper panels show how the total SFR enhancement varies with $t_{\rm peri}$, with 
the horizontal dashed lines denoting zero enhancement.
The left--hand, middle and right--hand columns correspond to eccentricities of 0.85, 0.95, and 1.05 respectively.
The colour scheme depicts impact parameters of 2 kpc (black), 4 kpc (blue), 
8 kpc (green), 12 kpc (magenta), and 16 kpc (red).
For clarity, we plot only the f disc orientation, which generally has SFR enhancements 
between the e and k disc orientations.
\label{figorbits}}
\end{figure}

We track the total and central (within 1 kpc) SFR of each simulated galaxy as a function of time.  
In order to facilitate comparison with our SDSS sample, we translate 
the simulated SFRs into SFR enhancements by normalizing 
by the SFR of the galaxy when evolved in isolation.  
In Fig.~\ref{figorbits}, we show how pair separation 
and total SFR enhancement depend on the time relative to first pericentre passage ($t_{\rm peri}$) 
for a representative subset of orbital configurations in this merger suite.  
The resulting $r_{\rm apo}$ varies from $\sim$55 to 220 kpc, 
with the largest $r_{\rm apo}$ corresponding to 
the highest eccentricity and largest impact parameter (and vice versa).  

For each of the orbits shown in Fig.~\ref{figorbits}, a strong burst of star formation 
begins shortly after the first close passage, with a maximum SFR enhancement of a factor of 1.3--2.2.  
This enhancement persists for about 1.5 Gyr for all orbits in which the galaxies remain apart 
from one another during this timeframe.  
In every case, a second and stronger burst of star formation begins shortly before coalescence.  
For orbits with relatively high eccentricities and impact parameters, 
there is enough time for the first episode of enhanced star formation to subside, followed 
by a period of suppressed star formation (quenching).  
This intriguing result would not have been seen if, like many other 
studies, we had restricted our orbits to low eccentricities and small impact parameters.

\subsection{SFR enhancements as a function of $r_p$}

\begin{figure}
\centerline{\rotatebox{0}{\resizebox{8.0cm}{!}
{\includegraphics{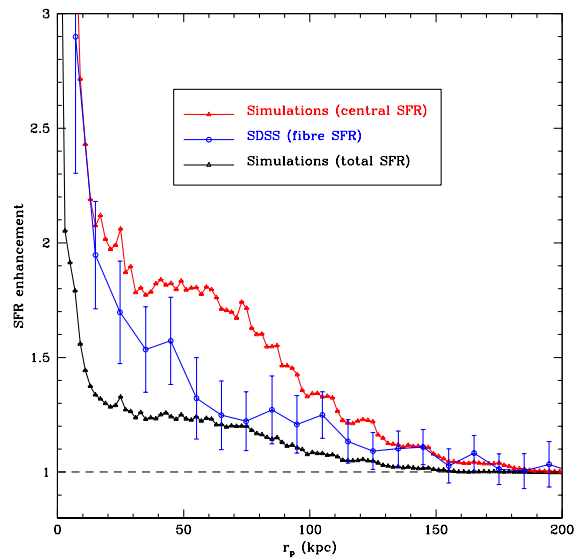}}}}
\caption{Mean SFR enhancement is plotted versus projected separation ($r_p$) for both 
the simulations (black and red symbols) and the observations (blue; from Fig. 1). 
The horizontal dashed line denotes zero SFR enhancement.  The SDSS SFRs are measured 
using fibres with a median projected radius of 2.5 kpc, whereas the central SFRs 
for the simulations are computed within a 3D radius of 1 kpc.
\label{figsimobs}}
\end{figure}

In order to compare these results with our SDSS measurements of SFR enhancements 
(Fig.~\ref{figrpsfr}), we observe each of these orbits from a large number of 
random orientations, and at random times during the orbits, computing $r_p$ and 
$\Delta v$ in each case.
While the simulated orbits begin at a initial separation of 165 kpc (corresponding to the 
virial radius of the more massive galaxy), we extrapolate each orbit backward in time to an initial 
separation of 10 Mpc, to ensure that we follow each orbit for effectively all the 
time that the pair has $r_p < 200$ kpc.  We then compute the average SFR enhancement 
over all 75 orbital configurations, treating each 10 Myr timestep equally, which accounts 
for the fact that some orbits take longer to coalesce than others.  We exclude all 
data points for which $\Delta v > 300$~\kms, as was done with our SDSS pairs.  The results are 
shown in Fig.~\ref{figsimobs} (the corresponding error bars 
are vanishingly small, since we average over so many viewing angles).  

Within the simulations, we see strong SFR enhancements at small separations ($r_p < 10$ kpc)  
which are driven by coalescing systems.  
The simulations yield moderate enhancements at a wide range 
of separations (roughly 20-100 kpc), with smaller enhancements extending out to $\sim$ 150 kpc.  
As seen in Fig.~\ref{figsimobs}, the total and central SFR enhancements in the simulations roughly bracket 
the SDSS enhancements, with a similar dependence on $r_p$.
These simulations suggest that the SFR enhancements seen in widely separated SDSS pairs 
are the result of star formation triggered by earlier close passages.  
In the simulations, it typically takes $\sim$ 1 Gyr for galaxies to 
reach $r_p \sim 150$ kpc 
after their first close passage, travelling at a time-averaged transverse speed of $\sim$ 150 \kms.  
This is consistent with the relatively long time-scale of enhanced star formation seen in 
Fig.~\ref{figorbits}.

\section{Discussion}

We have found clear evidence of enhanced star formation in pairs with separations as large 
as 150 kpc, and have demonstrated that such enhancements can be produced in simulations with 
realistic orbits.  
The rise in enhancements towards the smallest scales and the 
absence of any net enhancement on larger scales is consistent with star formation triggered 
by galaxy$-$galaxy interactions.  

Although it has been appreciated that mergers can drive starburst activity following 
close passages~\citep[e.g.,][]{BH96}, much of the focus has been on very close separations 
associated with final coalescence.  Here, we have used numerical simulations to show that 
this same physical picture can apply to SFR enhancements that occur 
at large galactic separations following first pericentre passage.  Specifically, since there is a 
time delay between close passages and peak starburst activity of the order of an orbital or 
dynamical time, it is possible for SFR enhancements to appear prominently 
at large galactic separations as the galaxies separate after first pericentre passage.

While a detailed census of interacting/merging pairs is beyond the scope of this Letter, 
we can estimate the fraction of star forming galaxies which are found in these wider pairs.  
Following \citet{patton00}, we weight the observed pair fraction by the reciprocal of the 
overall spectroscopic completeness of the survey (88 per cent according to \citealt{patton08}), 
and also weight close angular pairs 
by an additional factor of 3.08 (see Section~\ref{sdss}), and find that 12.3 per cent of galaxies are 
in $r_p < 150$ kpc pairs, versus 1.8 per cent in $r_p < 30$ kpc pairs.  
Moreover, by applying this methodology to the data in the lower panel of Fig.~\ref{figrpsfr}, 
we estimate that of the enhanced star formation occurring in pairs with $r_p < 150$ kpc, 
66 per cent occurs at $r_p > 30$ kpc.  This implies that interaction-induced star formation 
is much more prevalent than has been appreciated to date.  

We caution that our SDSS sample is not complete for the 1:10 stellar mass ratio pairs 
considered in this study, and we have not attempted to correct for this in our rough census.  
We note also that our SDSS paired galaxies (and their controls) were required to have secure SFR 
measurements; as a result, our sample is restricted primarily to `blue cloud' 
galaxies which lie on the star-forming sequence \citep{salim07}.  
Finally, we cannot rule out enhancements in individual pairs at even wider separations (including  
flybys); however, any such enhancements appear, to first order, to be cancelled out 
by the suppression of star formation in other pairs.  

We caution also that we have used a rather simple suite of merger simulations 
to demonstrate consistency between the observations and simulations.  
While this has the advantage of clarity, future studies will be able to 
improve on this 
by using a merger suite that more closely matches the properties of 
paired galaxies found in SDSS.  This can be achieved by 
using a substantially larger merger suite, allowing for coverage or exploration of a more 
complete portion of the merger parameter space.  Or, it may be possible to take advantage 
of continued advancements in computational resources and numerical 
methods to measure SFR enhancements directly  
from cosmological hydrodynamical
simulations.  Recent studies have shown improvements in the ability of galaxies to 
form and maintain gas--rich discs in 
cosmological simulations~\citep{Vogelsberger2012, torrey12b}
which may allow for direct measures of the SFR enhancement for 
paired galaxies using analogous methods to those applied in our SDSS analysis.

\section{Conclusions}

We have used a well-defined sample of $\sim$ 211 000 star-forming galaxies 
to measure SFR enhancements as a function of pair separation.  Our novel method of 
characterizing the local density and isolation of paired galaxies yields the first secure 
measurements of enhancements at wide separations.  
We find that SFR enhancements are detectable out to $r_p \sim 150$ kpc, 
with no net enhancement (or suppression) seen at larger separations.  
As with earlier work, we find the strongest enhancements at the smallest pair separations.
We have compared these results with the predictions from a suite of $N$--body/SPH simulations 
of merging galaxies, and find broad agreement.  In particular, the enhancements seen 
out to $\sim$ 150 kpc are a natural outcome of mergers in which the eccentricities and 
impact parameters are sufficiently high to lead to the required separations 
before the interaction--induced SFR enhancements subside.

\end{document}